\documentclass[aps,prd,onecolumn,groupedaddress,showpacs,nofootinbib,amssymb]{revtex4}
\usepackage[dvips]{graphicx}
\usepackage{amssymb}
\usepackage{amsmath}
\usepackage{graphicx,,color}
\usepackage{amsfonts}
\usepackage{bm}
\usepackage{cancel}
\usepackage{comment}

\newcommand\be{\begin{equation}}
\newcommand\ee{\end{equation}}

\allowdisplaybreaks[4]

\begin{document}

\tolerance=5000

\title{Swampland Implications of GW170817-compatible Einstein-Gauss-Bonnet Gravity}
\author{S.D.~Odintsov,$^{1,2}$\,\thanks{odintsov@ice.cat}
V.K.~Oikonomou,$^{3,4,5}$\,\thanks{v.k.oikonomou1979@gmail.com}}
\affiliation{$^{1)}$ ICREA, Passeig Luis Companys, 23, 08010 Barcelona, Spain\\
$^{2)}$ Institute of Space Sciences (IEEC-CSIC) C. Can Magrans
s/n,
08193 Barcelona, Spain\\
$^{3)}$ Department of Physics, Aristotle University of
Thessaloniki, Thessaloniki 54124,
Greece\\
$^{4)}$ Laboratory for Theoretical Cosmology, Tomsk State
University of Control Systems and Radioelectronics, 634050 Tomsk,
Russia (TUSUR)\\
$^{5)}$ Tomsk State Pedagogical University, 634061 Tomsk,
Russia\\}

\tolerance=5000

\begin{abstract}
We revisit the Einstein-Gauss-Bonnet theory in view of the
GW170817 event, which compels that the gravitational wave speed is
equal to $c_T^2=1$ in natural units. We use an alternative
approach compared to one previous work of ours, which enables us
to express all the slow-roll indices and the observational indices
as functions of the scalar field. Using our formalism we
investigate if the Swampland criteria are satisfied for the
Einstein-Gauss-Bonnet theory and as we demonstrate, the Swampland
criteria are satisfied for quite general forms of the potential
and the Gauss-Bonnet coupling function $\xi (\phi)$, if the
slow-roll conditions are assumed to hold true.
\end{abstract}


\maketitle

\section{Introduction}

The gravitational wave detection coming from the neutron star
merging GW170817 event \cite{GBM:2017lvd}, has utterly affected
modified theories of gravity, excluding some of these from being
viable descriptions of our Universe at astrophysical scales.
Particularly, the GW170817 event indicated that the propagating
speed of the gravitational waves $c_T$ is equal to unity, namely
$c_T^2=1$, in natural units. This feature as we mentioned,
excluded almost instantly many alternative theories of gravity,
see Ref. \cite{Ezquiaga:2017ekz} for a detailed account of this
topic. Nevertheless, there exist many modified gravity theories
that still remain robust against the GW170817 event results, see
for example the recent reviews
\cite{Nojiri:2017ncd,Nojiri:2010wj,Nojiri:2006ri,Capozziello:2011et,Capozziello:2010zz,delaCruzDombriz:2012xy,Olmo:2011uz}.

In a recent previous work of ours \cite{Odintsov:2019clh}, we
investigated how the Einstein-Gauss-Bonnet theories
\cite{Hwang:2005hb,Nojiri:2006je,Cognola:2006sp,Nojiri:2005vv,Nojiri:2005jg,Nojiri:2007te,Satoh:2008ck,Hikmawan:2015rze,Bamba:2014zoa,Yi:2018gse,Guo:2009uk,Guo:2010jr,Jiang:2013gza,Kanti:2015pda,vandeBruck:2017voa,Kanti:1998jd,Nozari:2017rta,Chakraborty:2018scm,Odintsov:2018zhw,Kawai:1998ab,Yi:2018dhl,vandeBruck:2016xvt,Kleihaus:2019rbg,Bakopoulos:2019tvc,Bakopoulos:2020dfg,Antoniadis:1993jc,Antoniadis:1990uu}
can be rendered viable and compatible with the GW170817 event. The
gravitational wave speed for an Einstein-Gauss-Bonnet theory is
equal to,
\begin{equation}
\label{GW} \centering c_T^2=1-\frac{Q_f}{2Q_t}\, ,
\end{equation}
with $Q_f=8c_1(\ddot\xi-H\dot\xi)$, while $c_1$ is a dimensionless
constant multiplication factor that enters the Lagrangian in front
of the Gauss-Bonnet term, and the function $Q_t$ is
$Q_t=\frac{1}{\kappa^2}-8c_1\dot\xi H$, with $H$ being the Hubble
rate. So according to our considerations, if $Q_f=0$ the
gravitational wave speed would be equal to one. This implies that
the Gauss-Bonnet coupling must satisfy the differential equation
$\ddot\xi-H\dot\xi=0$, which implies that $\dot{\xi}=e^{N}$, where
$N$ is the $e$-foldings number. This approach enabled us to
express all the slow-roll indices and the observational indices as
functions of the $e$-foldings number, and eventually study the
phenomenology of the model. In this letter we shall employ a
different approach, in order to express the slow-roll indices and
the observational indices as functions of the scalar field. Our
aim for this letter is to investigate whether the Swampland
criteria firstly obtained in Refs.
\cite{Vafa:2005ui,Ooguri:2006in} and later further studied in
Refs.
\cite{Palti:2020qlc,Brandenberger:2020oav,Blumenhagen:2019vgj,Wang:2019eym,Benetti:2019smr,Palti:2019pca,Cai:2018ebs,Mizuno:2019pcm,Aragam:2019khr,Brahma:2019mdd,Mukhopadhyay:2019cai,Marsh:2019lhu,Brahma:2019kch,Haque:2019prw,Heckman:2019dsj,Acharya:2018deu,Elizalde:2018dvw,Cheong:2018udx,Heckman:2018mxl,Kinney:2018nny,Garg:2018reu,Lin:2018rnx,Park:2018fuj,Olguin-Tejo:2018pfq,Fukuda:2018haz,Wang:2018kly,Ooguri:2018wrx,Matsui:2018xwa,Obied:2018sgi,Agrawal:2018own,Murayama:2018lie,Marsh:2018kub},
hold true, but as it will be apparent shortly, the slow-roll
indices and the corresponding observational indices have a very
simple functional form, thus the inflationary phenomenology is
considerably simplified in the present context, compared to our
previous approach \cite{Odintsov:2019clh}. An extensive account
for the inflationary phenomenology of Einstein-Gauss-Bonnet
theories using the formalism developed in the present letter, can
be found in \cite{PRDsubmitted}.

The result of this letter in short is that the Swampland criteria
for Einstein-Gauss-Bonnet theories of gravity can be satisfied for
general within assumptions scalar coupling function $\xi (\phi)$
and scalar field potential $V(\phi)$. Our result does not require
any specific functional relation between the scalar potential and
the scalar coupling function $\xi (\phi)$, apart from the fact
that in order for the condition (\ref{GW}) to be satisfied, the
potential $V(\phi)$ and the scalar coupling $\xi (\phi)$ must
satisfy a specific differential equation. Our result is somewhat
more general in comparison to the one of Ref. \cite{Yi:2018dhl},
where the Swampland criteria were addressed in the context of
Einstein-Gauss-Bonnet theories. As it was shown in
\cite{Yi:2018dhl}, if the function $\xi(\phi)$ is chosen as
$\xi(\phi )=\frac{\lambda}{V(\phi)}$, and also certain assumptions
are imposed on the scalar coupling function $\xi(\phi)$ and its
derivatives with respect to the cosmic time, the theory satisfies
the Swampland criteria. In our case, the constraint (\ref{GW})
does not require such restricted functional relations between
$\xi(\phi)$ and $V(\phi)$. In fact, if the relation $\xi(\phi
)=\frac{\lambda}{V(\phi)}$ holds true, as we show, this can also
be a subcase of solutions in the context of the present letter,
only if $\lambda=3/4$. Thus with this letter we provide a more
generic framework of Einstein-Gauss-Bonnet gravity, in the context
of which the Swampland criteria can be satisfied in a less
constrained way.

\section{The Swampland Criteria for the GW170817-compatible Einstein-Gauss-Bonnet Theory}

The Einstein-Gauss-Bonnet theory has the following gravitational
action,
\begin{equation}
\label{action} \centering
S=\int{d^4x\sqrt{-g}\left(\frac{R}{2\kappa^2}-\frac{1}{2}\omega\partial_\mu\phi\partial^\mu\phi-V(\phi)-\frac{1}{2}c_1\xi(\phi)\mathcal{G}\right)}\,
,
\end{equation}
with $\kappa=\frac{1}{M_p}$ and $M_P$ is the reduced Planck mass.
Also the function $\xi(\phi)$ is the Gauss-Bonnet scalar coupling,
which is dimensionless and $c_1$ is a dimensionless free variable,
which we shall set equal to unity hereafter, that is, $c_1=1$, for
simplicity. The parameter $\omega$ shall be assumed to be equal to
$\omega=1$, so that the kinetic term of the scalar field is
canonical, since the aim of this letter is to study the Swampland
criteria apply for well defined effective field theories, hence
the phantom case $\omega=-1$ is forbidden. Assuming a flat
Friedman-Robertson-Walker metric (FRW),
\begin{equation}
\label{metric} \centering
ds^2=-dt^2+a(t)^2\sum_{i=1}^{3}{(dx^{i})^2}\, ,
\end{equation}
the gravitational equations of motion read,
\begin{equation}
\label{motion1} \centering
\frac{3H^2}{\kappa^2}=\frac{1}{2}\omega\dot\phi^2+V+12 \dot\xi
H^3\, ,
\end{equation}
\begin{equation}
\label{motion2} \centering \frac{2\dot
H}{\kappa^2}=-\omega\dot\phi^2+4\ddot\xi H^2+8\dot\xi H\dot
H-4\dot\xi H^3\, ,
\end{equation}
\begin{equation}
\label{motion3} \centering \omega(\ddot\phi+3H\dot\phi)+V'+12
\xi'H^2(\dot H+H^2)=0\, .
\end{equation}
where the ``prime'' denotes differentiation with respect to the
scalar field $\phi$ hereafter.

The gravitational wave speed is given in Eq. (\ref{GW}), so by
requiring that this is equal to $c_T^2=1$, this implies that the
Gauss-Bonnet scalar coupling function must satisfy the
differential equation $\ddot\xi-H\dot\xi=0$ as we already
mentioned. By expressing the derivatives with respect to the
cosmic time, as $\frac{d}{dt}=\dot\phi\frac{d}{d\phi}$, we have
$\dot\xi=\xi'\dot\phi$ and we can write the differential equation
as,
\begin{equation}
\label{constraint1} \centering
\xi''\dot\phi^2+\xi'\ddot\phi=H\xi'\dot\phi\, .
\end{equation}
Assuming the slow-roll evolution holds true for the scalar field,
\begin{equation}\label{scalarfieldslowrollprev}
\frac{\dot\phi^2}{2} \ll V\, ,
\end{equation}
and by further assuming that the contribution of the second term
containing the term $\ddot{\phi}$ is subdominant, namely that,
$\xi'\ddot\phi \ll\xi''\dot\phi^2$, the equation
(\ref{constraint1}) becomes an algebraic equation which determines
$\dot{\phi}$ as follows,
\begin{equation}
\label{constraint} \centering
\dot{\phi}\simeq\frac{H\xi'}{\xi''}\, ,
\end{equation}
where recall that the ``prime'' denotes differentiation with
respect to the scalar field. In effect, by assuming that the
slow-roll condition $\dot H \ll H^2$ holds true in order for the
inflationary era to be realized in the first place, Eq.
(\ref{motion3}) can be re-written as follows,
\begin{equation}
\label{motion4} \centering
\frac{\xi'}{\xi''}\simeq-\frac{1}{3\omega H^2}\left(V'+12
\xi'H^4\right)\, .
\end{equation}
In addition, assuming that the following extra conditions
(\ref{scalarfieldslowroll}) hold true for the scalar field,
\begin{equation}\label{scalarfieldslowroll}
\ddot\phi\ll 3 H\dot\phi,\,\,\,12 \dot\xi H^3=12
\frac{\xi'^2H^4}{\xi''}\ll V\, ,
\end{equation}
we can rewrite the gravitational equations of motion Eq.
(\ref{motion1})-(\ref{motion3}) as follows,
\begin{equation}
\label{motion5} \centering H^2\simeq\frac{\kappa^2V}{3}\, ,
\end{equation}
\begin{equation}
\label{motion6} \centering \dot
H\simeq-\frac{1}{2}\kappa^2\omega\dot\phi^2\, ,
\end{equation}
\begin{equation}
\label{motion7} \centering \dot\phi\simeq-\frac{1}{3\omega
H}\left(V'+12 \xi'H^4\right)\, ,
\end{equation}
while recall that the constraint of Eq. (\ref{constraint1}) in
conjunction with the assumption $\xi'\ddot\phi
\ll\xi''\dot\phi^2$, results to the following condition,
\begin{equation}
\label{motion8} \centering \dot\phi\simeq\frac{H\xi'}{\xi''}\, .
\end{equation}
From Eqs. (\ref{motion7})  and (\ref{motion8}) we conclude that
the Gauss-Bonnet scalar coupling $\xi (\phi)$ must satisfy the
following differential equation,
\begin{equation}
\label{Vdifeq} \centering \frac{\xi'}{\xi''}=-\frac{1}{3\omega
H}\left( V'+12 \xi'H^4\right)\, ,
\end{equation}
or in view of Eq. (\ref{motion5}), we have the final form of the
differential equation that the Gauss-Bonnet scalar coupling $\xi
(\phi)$ must satisfy,
\begin{equation}
\label{maindiffeqn} \centering \frac{\xi'}{\xi''}=-\frac{1}{\omega
\kappa^2V}\Big{(}V' +\frac{4}{3} \xi'V^2 \kappa^4\Big{)}\, .
\end{equation}
Before proceeding, notice that the condition $12 \dot\xi H^3=12
\frac{\xi'^2H^4}{\xi''}\ll V$ appearing in Eq.
(\ref{scalarfieldslowroll}), in view of Eq. (\ref{motion5}) is
written as,
\begin{equation}\label{condition1EGNBswamp}
\Big{(}\frac{4}{3}  \kappa^4 \xi'V\Big{)}\frac{\xi'}{\xi''}\ll 1\,
,
\end{equation}
and notice that the above terms enter the differential equation
(\ref{maindiffeqn}). This will be a useful constraint when we
consider the Swampland criteria later on. The slow-roll indices
for an Einstein-Gauss-Bonnet theory are defined as follows
\cite{Hwang:2005hb},
\begin{align}\label{slowrollindicesdef}
\centering \epsilon_1&=-\frac{\dot H}{H^2},
&\epsilon_2&=\frac{\ddot\phi}{H\dot\phi}, & \epsilon_3&=0, &
\epsilon_4&=\frac{\dot E}{2HE}, & \epsilon_5&=\frac{Q_a}{2HQ_t},
&\epsilon_6&=\frac{\dot Q_t}{2HQ_t}\, ,
\end{align}
with the function $E$ being equal
to,
\begin{equation}\label{functionE}
E=\frac{1}{\kappa^2\dot\phi^2}\left(\omega\phi^2+3\left(\frac{Q_a^2}{2Q_t}\right)+Q_c\right)\,
,
\end{equation}
and the functions $Q_a$ and $Q_t$, $Q_b$ and $Q_c$, and also the
function $Q_e$ to be used later on, are defined as follows
\cite{Hwang:2005hb},
\begin{equation}\label{qis}
Q_a=-4 \dot\xi H^2,\,\,\,Q_b=-8 \dot\xi
H,\,\,\,Q_t=\frac{1}{\kappa^2}+\frac{Q_b}{2},\,\,\,Q_c=0,\,\,\,Q_e=-16
\dot{\xi}\dot{H}\, .
\end{equation}
In view of Eqs. (\ref{motion5})-(\ref{motion8}), we can express
the $Q_i$ functions of Eq. (\ref{qis}) as functions of the scalar
field. The $Q_i$ functions read,
\begin{equation}\label{Qa}
\centering Q_a\simeq-4 \frac{\xi'^2}{\xi''}H^3\simeq
-\frac{\left(4   \kappa ^3\right) V(\phi )^{3/2} \xi '(\phi
)^2}{\left(3 \sqrt{3}\right) \xi ''(\phi )}\, ,
\end{equation}
\begin{equation}\label{Qb}
\centering Q_b\simeq-8 \frac{\xi'^2}{\xi''}H^2\simeq
-\frac{\left(8   \kappa^2\right) V(\phi ) \xi '(\phi )^2}{3 \xi
''(\phi )}\, ,
\end{equation}
\begin{equation}\label{Qe}
\centering Q_e\simeq8
\kappa^2\omega\frac{\xi'^4}{\xi''^3}H^3\simeq \frac{V(\phi )^{3/2}
\left(8   \kappa ^5 \omega \right) \xi '(\phi )^4}{\left(3
\sqrt{3}\right) \xi ''(\phi )^3}\, .
\end{equation}
Also, the slow-roll indices of Eq. (\ref{slowrollindicesdef})
expressed in terms of the scalar field read,
\begin{equation}
\label{index1} \centering
\epsilon_1\simeq\frac{\kappa^2\omega}{2}\left(\frac{\xi'}{\xi''}\right)^2\,
,
\end{equation}
\begin{equation}
\label{index2} \centering
\epsilon_2\simeq-\epsilon_1+1-\frac{\xi'\xi'''}{\xi''^2}\, ,
\end{equation}
\begin{equation}
\label{index3} \centering \epsilon_3=0\, ,
\end{equation}
\begin{equation}
\label{index4} \centering
\epsilon_4\simeq\frac{\xi'}{2\xi''}\frac{E'}{E}\, ,
\end{equation}
\begin{equation}
\label{index5} \centering \epsilon_5\simeq-\frac{2\kappa^4
\xi'^2V}{3\xi''-4\kappa^4 \xi'^2V}\, ,
\end{equation}
\begin{equation}
\label{index6} \centering \epsilon_6\simeq-\frac{2\kappa^4
\xi'^2V\left(1-\frac{1}{2}\kappa^2\omega\left(\frac{\xi'}{\xi''}\right)^2\right)}{3\xi''-4
\kappa^4V\xi'^2}\, ,
\end{equation}
where the function $E(\phi)$ is equal to,
\begin{equation}\label{functionE}
E(\phi)=\frac{\omega}{\kappa^2}+\frac{8
^2\kappa^4\xi'^2V^2\xi''}{3\xi''\left(1-\frac{4\kappa^4
\xi'^2V}{3\xi''}\right)}\, .
\end{equation}
Also the spectral index of the primordial scalar curvature
perturbations $n_{\mathcal{S}}$, the spectral index of the tensor
perturbations $n_T$ and the tensor-to-scalar ratio $r$, as
functions of the scalar field $\phi$, are defined as follows
\cite{Hwang:2005hb},
\begin{equation}
\label{spectralindex} \centering
n_S=1-2\frac{2\epsilon_1+\epsilon_2+\epsilon_4}{1-\epsilon_1}\, ,
\end{equation}
\begin{equation}\label{tensorspectralindex}
n_T=-2\frac{\epsilon_1+\epsilon_6}{1-\epsilon_1}\, ,
\end{equation}
\begin{equation}\label{tensortoscalar}
r=16\left|\left(\frac{\kappa^2Q_e}{4H}-\epsilon_1\right)\frac{2c_A^3}{2+\kappa^2Q_b}\right|\,
,
\end{equation}
where $c_A$ is the speed of sound, which for the
Einstein-Gauss-Bonnet theory is equal to,
\begin{equation}
\label{sound} \centering
c_A^2=1+\frac{Q_aQ_e}{3Q_a^2+\omega\dot\phi^2(\frac{2}{\kappa^2}+Q_b)}\,
.
\end{equation}
It is apparent that the slow-roll indices, and correspondingly the
observational indices, have a very simple closed form, so this
will be valuable for inflationary phenomenology, the details of
which can be found in Ref. \cite{PRDsubmitted}. For completeness,
in order to further show the simplicity of the final equations
that determine the inflationary dynamics, we quote here the
relation of the $e$-foldings number $N$, defined as,
$N=\int_{t_i}^{t_f}{Hdt}$ (where $t_i$ and $t_f$ indicate the
initial and final time instance of inflation), which as a function
of the scalar field reads,
\begin{equation}
\label{efolds} \centering
N=\int_{\phi_i}^{\phi_f}{\frac{\xi''}{\xi'}d\phi}\, ,
\end{equation}
with $\phi_i$ and $\phi_f$ being the values of the scalar field at
the beginning and at the end of the inflationary era respectively.

Now let us proceed to the main focus of this paper, the Swampland
criteria \cite{Vafa:2005ui,Ooguri:2006in}. In the literature there
is currently a lot of activity regarding the implications of the
Swampland criteria, see Refs.
\cite{Palti:2020qlc,Brandenberger:2020oav,Blumenhagen:2019vgj,Wang:2019eym,Benetti:2019smr,Palti:2019pca,Cai:2018ebs,Mizuno:2019pcm,Aragam:2019khr,Brahma:2019mdd,Mukhopadhyay:2019cai,Marsh:2019lhu,Brahma:2019kch,Haque:2019prw,Heckman:2019dsj,Acharya:2018deu,Elizalde:2018dvw,Cheong:2018udx,Heckman:2018mxl,Lin:2018rnx,Park:2018fuj,Olguin-Tejo:2018pfq,Fukuda:2018haz,Wang:2018kly,Ooguri:2018wrx,Matsui:2018xwa,Obied:2018sgi,Agrawal:2018own,Murayama:2018lie,Marsh:2018kub}.
Also in Ref. \cite{Yi:2018dhl}, the Swampland criteria were
considered for Einstein-Gauss-Bonnet theories, like in our case,
but from a different perspective, since the compatibility with the
GW170817 event was not taken into account.

From now on we adopt the reduced Planck units physical system, so
that $\kappa^2=1$ and also take $\omega=1$ in order to have a
canonical kinetic term for the scalar field. The Swampland
criteria in reduced Planck units are the following,
\begin{enumerate}
    \item $\Delta \phi \leq d$, with $d\sim \mathcal{O}(1)$ in
    reduced Planck units.
    \item $\frac{V'}{V} \geq \gamma$, with $\gamma\sim \mathcal{O}(1)$ in
    reduced Planck units.
\end{enumerate}
The problem indicated by the Swampland criteria for a single
canonical scalar field is that the tensor to scalar ratio is
$r=16\left( \frac{V'}{V} \right)^2 $, so this must be larger than
unity, which is excluded by the latest 2018 Planck data.

Let us see if the Swampland criteria affect the
Einstein-Gauss-Bonnet inflationary phenomenology and how. From the
slow-roll condition $\frac{\dot\phi^2}{2} \ll V$ of Eq.
(\ref{scalarfieldslowroll}) for the scalar field, by substituting
$\dot{\phi}$ form Eq. (\ref{constraint}), we have,
\begin{equation}\label{basic1}
\frac{\frac{\dot\phi^2}{2}}{V}=\frac{\xi'^2H^2}{2\xi''^2V}=\frac{\xi'^2}{6\xi''^2}\ll
1\, ,
\end{equation}
in reduced Planck units. Defining $x=\frac{\xi'}{\sqrt{6}\xi''}$,
the condition (\ref{basic1}) becomes,
\begin{equation}\label{basic2}
x^2\ll 1\, .
\end{equation}
The tensor-to-scalar ratio (\ref{tensortoscalar}) is evaluated to
be equal to,
\begin{equation}\label{rexplicit}
r\simeq \Big{|}\frac{8 \xi '(\phi )^2 \left(\frac{-8 V(\phi )^2
\xi '(\phi )^2 \left(2 \xi '(\phi )^2-3 \xi ''(\phi )^2\right)-12
V(\phi ) \xi '(\phi )^2 \xi ''(\phi )+9 \xi ''(\phi )^2}{\xi
''(\phi ) \left(8  V(\phi )^2 \xi '(\phi )^2 \xi ''(\phi )-4
V(\phi ) \xi '(\phi )^2+3 \xi ''(\phi )\right)}\right)^{3/2}}{3
\sqrt{3} \xi ''(\phi )^2} \Big{|}\, ,
\end{equation}
for general $x(\phi)$ and $V(\phi)$. Now since $\frac{V'}{V}\sim
\gamma $, we substitute this in the relation (\ref{maindiffeqn}),
so we obtain that $\xi''$ is equal to,
\begin{equation}\label{xddot}
\xi''=\frac{\xi '(\phi )}{-\gamma -\frac{4}{3} V(\phi ) \xi '(\phi
)}\, ,
\end{equation}
and also by using the definition of $x$, by using Eq.
(\ref{maindiffeqn}), it can easily be shown that $\xi'$ can be
expressed in terms of the potential $V$, $x$ and of the function
$\gamma$ as follows,
\begin{equation}\label{xiprime}
\xi'=\frac{3}{4}\frac{\sqrt{6}x-\gamma}{V}\, .
\end{equation}
Substituting this in Eq. (\ref{rexplicit}) we can obtain an
expression for the tensor-to-scalar ratio as a function of the
function $\gamma$, so the final expression is,
\begin{equation}\label{rtensorsemifinal}
r=\Big{|}\frac{16 x^2 \left(\frac{-9 \left(4 x^2-1\right)
\left(\gamma ^2+6 x^2-2 \sqrt{6} \gamma  x\right)+6 x \left(6
x-\sqrt{6} \gamma \right)+6}{3 \left(\gamma^2+6 x^2-2 \sqrt{6}
\gamma  x\right)+2 x \left(6 x-\sqrt{6} \gamma
\right)+2}\right)^{3/2}}{\sqrt{3}}\Big{|}\, .
\end{equation}
Now by replacing $x=\sqrt{\varepsilon}$ in the above expression,
we have,
\begin{equation}\label{rsemifinal2}
r\simeq \Big{|}-48 \varepsilon  \left(\frac{24 \sqrt{6}
\varepsilon ^{3/2}-72 \varepsilon^2+18 \varepsilon -8 \sqrt{6}
\sqrt{\varepsilon }+5}{30 \varepsilon -8 \sqrt{6}
\sqrt{\varepsilon }+5}\right)^{3/2}\Big{|}\, ,
\end{equation}
and since $x^2=\varepsilon \ll 1$, by taking the Taylor expansion
of the expression (\ref{rsemifinal2}) for $\varepsilon\to 0$, we
get,
\begin{equation}\label{rfinal}
r\simeq \Big{|}-48 \varepsilon+\frac{864 \gamma^2 \varepsilon
^2}{3 \gamma ^2+2}+\mathcal{O}(\varepsilon^{5/2})+...\Big{|}\, ,
\end{equation}
at leading order. So by using again $x=\sqrt{\varepsilon}$, we
have that the tensor-to-scalar ratio at leading order reads,
\begin{equation}\label{tensortoscalarfinalnewref}
r\simeq 48 x^2\, ,
\end{equation}
so in view of Eq. (\ref{basic2}), we have that $r\ll 1$. Also,
$\gamma\sim \mathcal{O}(1)$, so it does not affect the Taylor
expansion. Therefore, we demonstrated that the Swampland criteria
do not affect the inflationary phenomenology of the
Einstein-Gauss-Bonnet theory of gravity, when the latter is
required to be compatible with the GW170817 event.

It is easy to show that the same result can be obtained, by using
an alternative way, if the following condition holds true in
reduced Planck units,
\begin{equation}\label{alt1}
\frac{\xi'}{\sqrt{6}\xi''}\ll 1\, .
\end{equation}
Recall that the fraction $\frac{\xi'}{\xi''}$ enters in the
condition appearing in Eq. (\ref{condition1EGNBswamp}), but more
importantly in the differential equation (\ref{maindiffeqn}). In
view of the new condition (\ref{alt1}), the differential equation
(\ref{maindiffeqn}) becomes,
\begin{equation}
\label{maindiffeqnalt1} \centering \frac{V'}{V} +\frac{4}{3}\xi'V
=0\, ,
\end{equation}
where we took $\omega=\kappa^2=1$. So in this case, by
substituting $\xi ''\to \frac{\xi '}{\sqrt{6} x}$ in
(\ref{rtensorsemifinal}) and subsequently by substituting $\xi'\to
\frac{3 \gamma }{4 V(\phi )}$ which is obtained from Eq.
(\ref{maindiffeqnalt1}), we obtain,
\begin{equation}\label{ralternative3}
r\simeq \Big{|}-48 \varepsilon +\frac{864 \gamma ^2 \varepsilon
^2}{3 \gamma ^2+2}+\mathcal{O}(\varepsilon^{5/2})\Big{|}\, ,
\end{equation}
which is identical to the one appearing in Eq. (\ref{rfinal}),
with the difference that the result of Eq. (\ref{rfinal}) was
obtained without assuming the extra condition (\ref{alt1}), namely
$\frac{\xi'}{\sqrt{6}\xi''}\ll 1$.

In fact, this extra condition $\frac{\xi'}{\sqrt{6}\xi''}\ll 1$
can be related to any phenomenologically viable
Einstein-Gauss-Bonnet theory compatible with the GW170817 event,
and as we show in Ref. \cite{PRDsubmitted}, it proves that the
condition $\frac{\xi'}{\sqrt{6}\xi''}\ll 1$ simplifies
significantly the calculations but more importantly, it seems to
be an inherent characteristic of the inflationary phenomenology of
any viable Einstein-Gauss-Bonnet theory compatible with the
GW170817 event.

Let us discuss here in brief the results of the paper
\cite{Yi:2018dhl}, since some solutions may have some
phenomenological importance. As it was shown in \cite{Yi:2018dhl},
if the function $\xi(\phi)$ is chosen as $\xi(\phi
)=\frac{\lambda}{V(\phi)}$, where $V(\phi)$ is the scalar
potential appearing in the action (\ref{action}). However, if the
compatibility with the GW170817 is imposed on the
Einstein-Gauss-Bonnet theory is imposed, the function $\xi(\phi
)=\frac{\lambda}{V(\phi)}$ does not satisfy in general the
differential equation (\ref{maindiffeqn}), unless the potential
has a very specific form, which however does not yield a viable
inflationary phenomenology in the context of the present paper
\cite{PRDsubmitted}. However, if the condition
$\frac{\xi'}{\xi''}\ll 1$ is imposed in the theory, then the
function $\xi(\phi )=\frac{\lambda}{V(\phi)}$ satisfies the
differential equation the differential equation
(\ref{maindiffeqnalt1}), only if $\lambda=3/4$. In Ref.
\cite{PRDsubmitted} we also address the phenomenology of this
class of models.

Another issue that is worth commenting on is that the authors of
Refs. \cite{Yi:2018gse,Yi:2018dhl} used a smaller number of
slow-roll indices, in comparison to our approach. This is possibly
due to the fact that one of the assumptions they used in order to
calculate the power spectrum was $\ddot{\xi}\ll \dot{\xi}H$, which
in contrast we assumed that $\ddot{\xi}=H\dot{\xi}$, see above Eq.
(\ref{constraint1}). The assumption $\ddot{\xi}\ll \dot{\xi}H$
lead to simpler expressions of the power spectrum, however, the
gravitational wave speed in their context is $c_T^2\neq 1$.

Finally, let us consider here the Lyth bound \cite{Lyth:1996im}
implications in the context of the GW170817-corrected
Einstein-Gauss-Bonnet theory. Recall that the Lyth bound is
\cite{Lyth:1996im} $\Delta \phi>\Delta N \sqrt{\frac{r}{8}}$, and
also the first Swampland criterion constrains $\Delta \phi$ to be
$\Delta \phi \leq d$, with $d\sim \mathcal{O}(1)$ in reduced
Planck units. Thus we have the constraint $d \geq \Delta
\phi>\Delta N \sqrt{\frac{r}{8}}$. In our case, $r\simeq
48x^2=\frac{8 \xi '(\phi )^2}{\xi ''(\phi )^2}$, thus the
constraint on $\Delta \phi $ becomes $d \geq \Delta \phi>\Delta N
\frac{ \xi '}{\xi ''}$. In effect, the condition $\Delta
\phi>\Delta N \frac{ \xi '}{\xi ''}$ constraints further the
model, so for example if $\Delta N\sim 60$, in order for the Lyth
bound to be satisfied, a viable model that is constrained to
satisfy the Swampland criteria, must further satisfy $\Delta
\phi>60 \frac{ \xi '}{\xi ''}$. The present framework used in this
letter guarantees that $x^2\ll 1$, where recall that $x$ defined
in the previous text is $x=\frac{\xi'}{\sqrt{6}\xi''}$, however,
the Lyth bound further constrains the Swampland compatible viable
inflationary models, which now must satisfy additionally $\Delta
\phi>60 \frac{ \xi '}{\xi ''}$. If for example $\frac{ \xi '}{\xi
''}\sim \mathcal{O}(10^{-2})$, then we would have $d \geq \Delta
\phi> 0.6$, so indeed the lower bound imposed by the Lyth bound,
is smaller than $d$ which is of the order of unity in reduced
Planck units. Also, if $\frac{ \xi '}{\xi ''}\sim
\mathcal{O}(10^{-2})$, then $x^2\sim \mathcal{O}(1.66\times
10^{-5})$, which is also compatible with the earlier assumed
constraint (\ref{basic2}), that is $x^2\ll 1$. Thus the Lyth bound
imposes the additional constraint $d \geq \Delta \phi>\Delta N
\frac{ \xi '}{\xi ''}$ in the theory, and should certainly be
taken into account in order to construct a phenomenologically
viable theory that also respects the Swampland criteria.

\section{Conclusions}

In this letter we revisited the Einstein-Gauss-Bonnet theory in
view of the GW170817 event, which compelled that the gravitational
wave speed is equal to $c_T^2\simeq 1$ in natural units. Using the
formalism we developed for the GW170817-compatible
Einstein-Gauss-Bonnet theory, we investigated if the Swampland
criteria are satisfied for the theory at hand, and as we
demonstrated, the Swampland criteria are satisfied naturally and
for quite general forms of the scalar potential and of the
Gauss-Bonnet scalar coupling function $\xi (\phi)$, assuming that
the slow-roll conditions for the scalar field and for the Hubble
rate hold true.

An interesting future perspective of the current work, again in
the context of the Swampland criteria, is that in the same way,
one can in principle consider superstring cosmology
\cite{Maeda:2011zn,Easther:1996yd} in the presence of two scalars
in string spectrum, thus not just considering simply the dilaton
field, as in the present work.

\section*{Acknowledgments}

This work is supported by MINECO (Spain), FIS2016-76363-P, and by
project 2017 SGR247 (AGAUR, Catalonia) (S.D.O).


\begin{thebibliography}{99}



\bibitem{GBM:2017lvd}
  B.~P.~Abbott {\it et al.}
  ``Multi-messenger Observations of a Binary Neutron Star Merger,''
  Astrophys.\ J.\  {\bf 848} (2017) no.2,  L12
  doi:10.3847/2041-8213/aa91c9
  [arXiv:1710.05833 [astro-ph.HE]].




\bibitem{Ezquiaga:2017ekz}
  J.~M.~Ezquiaga and M.~Zumalacarregui,
  Phys.\ Rev.\ Lett.\  {\bf 119} (2017) no.25,  251304
  doi:10.1103/PhysRevLett.119.251304
  [arXiv:1710.05901 [astro-ph.CO]].




\bibitem{Nojiri:2017ncd}
S.~Nojiri, S.~D.~Odintsov and V.~K.~Oikonomou,
Phys.\ Rept.\ {\bf 692} (2017) 1 doi:10.1016/j.physrep.2017.06.001
[arXiv:1705.11098 [gr-qc]].

\bibitem{Nojiri:2010wj}
S.~Nojiri and S.~D.~Odintsov,
Phys.\ Rept.\ {\bf 505} (2011) 59
doi:10.1016/j.physrep.2011.04.001 [arXiv:1011.0544 [gr-qc]].

\bibitem{Nojiri:2006ri}
S.~Nojiri and S.~D.~Odintsov,
eConf C {\bf 0602061} (2006) 06
 [Int.\ J.\ Geom.\ Meth.\ Mod.\ Phys.\ {\bf 4} (2007) 115]
doi:10.1142/S0219887807001928 [hep-th/0601213].

\bibitem{Capozziello:2011et}
S.~Capozziello and M.~De Laurentis,
Phys.\ Rept.\ {\bf 509} (2011) 167
doi:10.1016/j.physrep.2011.09.003 [arXiv:1108.6266 [gr-qc]].

\bibitem{Capozziello:2010zz}
V.~Faraoni and S.~Capozziello,
Fundam.\ Theor.\ Phys.\ {\bf 170} (2010).
doi:10.1007/978-94-007-0165-6

\bibitem{delaCruzDombriz:2012xy}
A.~de la Cruz-Dombriz and D.~Saez-Gomez,
Entropy {\bf 14} (2012) 1717 doi:10.3390/e14091717
[arXiv:1207.2663 [gr-qc]].

\bibitem{Olmo:2011uz}
G.~J.~Olmo,
Int.\ J.\ Mod.\ Phys.\ D {\bf 20} (2011) 413
doi:10.1142/S0218271811018925 [arXiv:1101.3864 [gr-qc]].


\bibitem{Odintsov:2019clh}
  S.~D.~Odintsov and V.~K.~Oikonomou,
  Phys.\ Lett.\ B {\bf 797} (2019) 134874
  doi:10.1016/j.physletb.2019.134874
  [arXiv:1908.07555 [gr-qc]].





\bibitem{Hwang:2005hb}
  J.~c.~Hwang and H.~Noh,
  Phys.\ Rev.\ D {\bf 71} (2005) 063536
  doi:10.1103/PhysRevD.71.063536
  [gr-qc/0412126].


\bibitem{Nojiri:2006je}
  S.~Nojiri, S.~D.~Odintsov and M.~Sami,
  Phys.\ Rev.\ D {\bf 74} (2006) 046004
  doi:10.1103/PhysRevD.74.046004
  [hep-th/0605039].




\bibitem{Cognola:2006sp}
  G.~Cognola, E.~Elizalde, S.~Nojiri, S.~Odintsov and S.~Zerbini,
  Phys.\ Rev.\ D {\bf 75} (2007) 086002
  doi:10.1103/PhysRevD.75.086002
  [hep-th/0611198].



\bibitem{Nojiri:2005vv}
  S.~Nojiri, S.~D.~Odintsov and M.~Sasaki,
  Phys.\ Rev.\ D {\bf 71} (2005) 123509
  doi:10.1103/PhysRevD.71.123509
  [hep-th/0504052].


\bibitem{Nojiri:2005jg}
  S.~Nojiri and S.~D.~Odintsov,
  Phys.\ Lett.\ B {\bf 631} (2005) 1
  doi:10.1016/j.physletb.2005.10.010
  [hep-th/0508049].


\bibitem{Nojiri:2007te}
  S.~Nojiri, S.~D.~Odintsov and P.~V.~Tretyakov,
  Phys.\ Lett.\ B {\bf 651} (2007) 224
  doi:10.1016/j.physletb.2007.06.029
  [arXiv:0704.2520 [hep-th]].


\bibitem{Satoh:2008ck}
  M.~Satoh and J.~Soda,
  JCAP {\bf 0809} (2008) 019
  doi:10.1088/1475-7516/2008/09/019
  [arXiv:0806.4594 [astro-ph]].




\bibitem{Hikmawan:2015rze}
  G.~Hikmawan, J.~Soda, A.~Suroso and F.~P.~Zen,
  Phys.\ Rev.\ D {\bf 93} (2016) no.6,  068301
  doi:10.1103/PhysRevD.93.068301
  [arXiv:1512.00222 [hep-th]].



\bibitem{Bamba:2014zoa}
  K.~Bamba, A.~N.~Makarenko, A.~N.~Myagky and S.~D.~Odintsov,
  JCAP {\bf 1504} (2015) 001
  doi:10.1088/1475-7516/2015/04/001
  [arXiv:1411.3852 [hep-th]].


\bibitem{Yi:2018gse}
  Z.~Yi, Y.~Gong and M.~Sabir,
  Phys.\ Rev.\ D {\bf 98} (2018) no.8,  083521
  doi:10.1103/PhysRevD.98.083521
  [arXiv:1804.09116 [gr-qc]].


\bibitem{Guo:2009uk}
  Z.~K.~Guo and D.~J.~Schwarz,
  Phys.\ Rev.\ D {\bf 80} (2009) 063523
  doi:10.1103/PhysRevD.80.063523
  [arXiv:0907.0427 [hep-th]].


\bibitem{Guo:2010jr}
  Z.~K.~Guo and D.~J.~Schwarz,
  Phys.\ Rev.\ D {\bf 81} (2010) 123520
  doi:10.1103/PhysRevD.81.123520
  [arXiv:1001.1897 [hep-th]].


\bibitem{Jiang:2013gza}
  P.~X.~Jiang, J.~W.~Hu and Z.~K.~Guo,
  Phys.\ Rev.\ D {\bf 88} (2013) 123508
  doi:10.1103/PhysRevD.88.123508
  [arXiv:1310.5579 [hep-th]].



\bibitem{Kanti:2015pda}
  P.~Kanti, R.~Gannouji and N.~Dadhich,
  Phys.\ Rev.\ D {\bf 92} (2015) no.4,  041302
  doi:10.1103/PhysRevD.92.041302
  [arXiv:1503.01579 [hep-th]].


\bibitem{vandeBruck:2017voa}
  C.~van de Bruck, K.~Dimopoulos, C.~Longden and C.~Owen,
  arXiv:1707.06839 [astro-ph.CO].



\bibitem{Kanti:1998jd}
  P.~Kanti, J.~Rizos and K.~Tamvakis,
  Phys.\ Rev.\ D {\bf 59} (1999) 083512
  doi:10.1103/PhysRevD.59.083512
  [gr-qc/9806085].





\bibitem{Nozari:2017rta}
  K.~Nozari and N.~Rashidi,
  Phys.\ Rev.\ D {\bf 95} (2017) no.12,  123518
  doi:10.1103/PhysRevD.95.123518
  [arXiv:1705.02617 [astro-ph.CO]].



\bibitem{Chakraborty:2018scm}
  S.~Chakraborty, T.~Paul and S.~SenGupta,
  Phys.\ Rev.\ D {\bf 98} (2018) no.8,  083539
  doi:10.1103/PhysRevD.98.083539
  [arXiv:1804.03004 [gr-qc]].



\bibitem{Odintsov:2018zhw}
  S.~D.~Odintsov and V.~K.~Oikonomou,
  Phys.\ Rev.\ D {\bf 98} (2018) no.4,  044039
  doi:10.1103/PhysRevD.98.044039
  [arXiv:1808.05045 [gr-qc]].


  \bibitem{Kawai:1998ab}
  S.~Kawai, M.~a.~Sakagami and J.~Soda,
  Phys.\ Lett.\ B {\bf 437}, 284 (1998)
  doi:10.1016/S0370-2693(98)00925-3
  [gr-qc/9802033].


\bibitem{Yi:2018dhl}
  Z.~Yi and Y.~Gong,
  Universe {\bf 5} (2019) no.9,  200
  doi:10.3390/universe5090200
  [arXiv:1811.01625 [gr-qc]].


\bibitem{vandeBruck:2016xvt}
  C.~van de Bruck, K.~Dimopoulos and C.~Longden,
  Phys.\ Rev.\ D {\bf 94} (2016) no.2,  023506
  doi:10.1103/PhysRevD.94.023506
  [arXiv:1605.06350 [astro-ph.CO]].


\bibitem{Kleihaus:2019rbg}
  B.~Kleihaus, J.~Kunz and P.~Kanti,
  arXiv:1910.02121 [gr-qc].





\bibitem{Bakopoulos:2019tvc}
  A.~Bakopoulos, P.~Kanti and N.~Pappas,
  Phys.\ Rev.\ D {\bf 101} (2020) no.4,  044026
  doi:10.1103/PhysRevD.101.044026
  [arXiv:1910.14637 [hep-th]].


\bibitem{Bakopoulos:2020dfg}
  A.~Bakopoulos, P.~Kanti and N.~Pappas,
  arXiv:2003.02473 [hep-th].


\bibitem{Antoniadis:1993jc}
  I.~Antoniadis, J.~Rizos and K.~Tamvakis,
  Nucl.\ Phys.\ B {\bf 415} (1994) 497
  doi:10.1016/0550-3213(94)90120-1
  [hep-th/9305025].

\bibitem{Antoniadis:1990uu}
I.~Antoniadis, C.~Bachas, J.~R.~Ellis and D.~V.~Nanopoulos,
Phys.\ Lett.\ B \textbf{257} (1991), 278-284
doi:10.1016/0370-2693(91)91893-Z














\bibitem{Vafa:2005ui}
  C.~Vafa,
  hep-th/0509212.


\bibitem{Ooguri:2006in}
  H.~Ooguri and C.~Vafa,
  Nucl.\ Phys.\ B {\bf 766} (2007) 21
  doi:10.1016/j.nuclphysb.2006.10.033
  [hep-th/0605264].



\bibitem{Palti:2020qlc}
  E.~Palti, C.~Vafa and T.~Weigand,
  arXiv:2003.10452 [hep-th].





\bibitem{Brandenberger:2020oav}
  R.~Brandenberger, V.~Kamali and R.~O.~Ramos,
  arXiv:2002.04925 [hep-th].



\bibitem{Blumenhagen:2019vgj}
  R.~Blumenhagen, M.~Brinkmann and A.~Makridou,
  JHEP {\bf 2002} (2020) 064
   [JHEP {\bf 2020} (2020) 064]
  doi:10.1007/JHEP02(2020)064
  [arXiv:1910.10185 [hep-th]].



\bibitem{Wang:2019eym}
  Z.~Wang, R.~Brandenberger and L.~Heisenberg,
  arXiv:1907.08943 [hep-th].





\bibitem{Benetti:2019smr}
  M.~Benetti, S.~Capozziello and L.~L.~Graef,
  Phys.\ Rev.\ D {\bf 100} (2019) no.8,  084013
  doi:10.1103/PhysRevD.100.084013
  [arXiv:1905.05654 [gr-qc]].




\bibitem{Palti:2019pca}
  E.~Palti,
  Fortsch.\ Phys.\  {\bf 67} (2019) no.6,  1900037
  doi:10.1002/prop.201900037
  [arXiv:1903.06239 [hep-th]].




\bibitem{Cai:2018ebs}
  R.~G.~Cai, S.~Khimphun, B.~H.~Lee, S.~Sun, G.~Tumurtushaa and Y.~L.~Zhang,
  Phys.\ Dark Univ.\  {\bf 26} (2019) 100387
  doi:10.1016/j.dark.2019.100387
  [arXiv:1812.11105 [hep-th]].

\bibitem{Mizuno:2019pcm}
  S.~Mizuno, S.~Mukohyama, S.~Pi and Y.~L.~Zhang,
  JCAP {\bf 1909} (2019) no.09,  072
  doi:10.1088/1475-7516/2019/09/072
  [arXiv:1905.10950 [hep-th]].




\bibitem{Aragam:2019khr}
  V.~Aragam, S.~Paban and R.~Rosati,
  arXiv:1905.07495 [hep-th].





\bibitem{Brahma:2019mdd}
  S.~Brahma and M.~W.~Hossain,
  Phys.\ Rev.\ D {\bf 100} (2019) no.8,  086017
  doi:10.1103/PhysRevD.100.086017
  [arXiv:1904.05810 [hep-th]].




\bibitem{Mukhopadhyay:2019cai}
  U.~Mukhopadhyay and D.~Majumdar,
  Phys.\ Rev.\ D {\bf 100} (2019) no.2,  024006
  doi:10.1103/PhysRevD.100.024006
  [arXiv:1904.01455 [gr-qc]].




\bibitem{Marsh:2019lhu}
  D.~M.~C.~Marsh and J.~E.~D.~Marsh,
  arXiv:1903.12643 [hep-th].




\bibitem{Brahma:2019kch}
  S.~Brahma and M.~W.~Hossain,
  JHEP {\bf 1906} (2019) 070
  doi:10.1007/JHEP06(2019)070
  [arXiv:1902.11014 [hep-th]].



\bibitem{Haque:2019prw}
  M.~R.~Haque and D.~Maity,
  Phys.\ Rev.\ D {\bf 99} (2019) no.10,  103534
  doi:10.1103/PhysRevD.99.103534
  [arXiv:1902.09491 [hep-th]].

\bibitem{Heckman:2019dsj}
  J.~J.~Heckman, C.~Lawrie, L.~Lin, J.~Sakstein and G.~Zoccarato,
  Fortsch.\ Phys.\  {\bf 67} (2019) no.11,  1900071
  doi:10.1002/prop.201900071
  [arXiv:1901.10489 [hep-th]].

\bibitem{Acharya:2018deu}
  B.~S.~Acharya, A.~Maharana and F.~Muia,
  JHEP {\bf 1903} (2019) 048
  doi:10.1007/JHEP03(2019)048
  [arXiv:1811.10633 [hep-th]].

\bibitem{Elizalde:2018dvw}
E.~Elizalde and M.~Khurshudyan,
Phys.\ Rev.\ D \textbf{99} (2019) no.10, 103533
doi:10.1103/PhysRevD.99.103533 [arXiv:1811.03861 [astro-ph.CO]].


\bibitem{Cheong:2018udx}
  D.~Y.~Cheong, S.~M.~Lee and S.~C.~Park,
  Phys.\ Lett.\ B {\bf 789} (2019) 336
  doi:10.1016/j.physletb.2018.12.046
  [arXiv:1811.03622 [hep-ph]].

\bibitem{Heckman:2018mxl}
  J.~J.~Heckman, C.~Lawrie, L.~Lin and G.~Zoccarato,
  Fortsch.\ Phys.\  {\bf 67} (2019) no.10,  1900057
  doi:10.1002/prop.201900057
  [arXiv:1811.01959 [hep-th]].


\bibitem{Kinney:2018nny}
W.~H.~Kinney, S.~Vagnozzi and L.~Visinelli,
Class.\ Quant.\ Grav.\  \textbf{36} (2019) no.11, 117001
doi:10.1088/1361-6382/ab1d87 [arXiv:1808.06424 [astro-ph.CO]].



\bibitem{Garg:2018reu}
  S.~K.~Garg and C.~Krishnan,
  JHEP {\bf 1911} (2019) 075
  doi:10.1007/JHEP11(2019)075
  [arXiv:1807.05193 [hep-th]].

\bibitem{Lin:2018rnx}
  C.~M.~Lin,
  Phys.\ Rev.\ D {\bf 99} (2019) no.2,  023519
  doi:10.1103/PhysRevD.99.023519
  [arXiv:1810.11992 [astro-ph.CO]].

\bibitem{Park:2018fuj}
  S.~C.~Park,
  JCAP {\bf 1901} (2019) 053
  doi:10.1088/1475-7516/2019/01/053
  [arXiv:1810.11279 [hep-ph]].

\bibitem{Olguin-Tejo:2018pfq}
  Y.~Olguin-Trejo, S.~L.~Parameswaran, G.~Tasinato and I.~Zavala,
  JCAP {\bf 1901} (2019) 031
  doi:10.1088/1475-7516/2019/01/031
  [arXiv:1810.08634 [hep-th]].





\bibitem{Fukuda:2018haz}
  H.~Fukuda, R.~Saito, S.~Shirai and M.~Yamazaki,
  Phys.\ Rev.\ D {\bf 99} (2019) no.8,  083520
  doi:10.1103/PhysRevD.99.083520
  [arXiv:1810.06532 [hep-th]].


\bibitem{Wang:2018kly}
  S.~J.~Wang,
  Phys.\ Rev.\ D {\bf 99} (2019) no.2,  023529
  doi:10.1103/PhysRevD.99.023529
  [arXiv:1810.06445 [hep-th]].


\bibitem{Ooguri:2018wrx}
  H.~Ooguri, E.~Palti, G.~Shiu and C.~Vafa,
  Phys.\ Lett.\ B {\bf 788} (2019) 180
  doi:10.1016/j.physletb.2018.11.018
  [arXiv:1810.05506 [hep-th]].

\bibitem{Matsui:2018xwa}
  H.~Matsui, F.~Takahashi and M.~Yamada,
  Phys.\ Lett.\ B {\bf 789} (2019) 387
  doi:10.1016/j.physletb.2018.12.055
  [arXiv:1809.07286 [astro-ph.CO]].





\bibitem{Obied:2018sgi}
  G.~Obied, H.~Ooguri, L.~Spodyneiko and C.~Vafa,
  arXiv:1806.08362 [hep-th].


\bibitem{Agrawal:2018own}
  P.~Agrawal, G.~Obied, P.~J.~Steinhardt and C.~Vafa,
  Phys.\ Lett.\ B {\bf 784} (2018) 271
  doi:10.1016/j.physletb.2018.07.040
  [arXiv:1806.09718 [hep-th]].






\bibitem{Murayama:2018lie}
  H.~Murayama, M.~Yamazaki and T.~T.~Yanagida,
  JHEP {\bf 1812} (2018) 032
  doi:10.1007/JHEP12(2018)032
  [arXiv:1809.00478 [hep-th]].

\bibitem{Marsh:2018kub}
  M.~C.~David Marsh,
  Phys.\ Lett.\ B {\bf 789} (2019) 639
  doi:10.1016/j.physletb.2018.11.001
  [arXiv:1809.00726 [hep-th]].


  \bibitem{PRDsubmitted} S.D. Odintsov, V.K. Oikonomou, F.P.
  Fronimos, ``Rectifying Einstein-Gauss-Bonnet Inflation in View of
  GW170817'', Nuclear Physics B submitted. [arXiv:2003.13724 [gr-qc]]


\bibitem{Lyth:1996im}
  D.~H.~Lyth,
  Phys.\ Rev.\ Lett.\  {\bf 78} (1997) 1861
  doi:10.1103/PhysRevLett.78.1861
  [hep-ph/9606387].




\bibitem{Maeda:2011zn}
  K.~i.~Maeda, N.~Ohta and R.~Wakebe,
  Eur.\ Phys.\ J.\ C {\bf 72} (2012) 1949
  doi:10.1140/epjc/s10052-012-1949-6
  [arXiv:1111.3251 [hep-th]].



\bibitem{Easther:1996yd}
  R.~Easther and K.~i.~Maeda,
  Phys.\ Rev.\ D {\bf 54} (1996) 7252
  doi:10.1103/PhysRevD.54.7252
  [hep-th/9605173].



\end{thebibliography}
\end{document}